\documentclass[twocolumn,pre,aps,showpacs,amsmath,amssymb,superscriptaddress]{revtex4}

\usepackage{hyperref}
\usepackage{amsmath,amssymb}
\usepackage{amsfonts,amsthm}
\usepackage{graphics}
\usepackage{graphicx}
\usepackage{dcolumn}
\usepackage{color}
\usepackage{bm}

\usepackage[normalem]{ulem}

\begin{document}


\title{ Inhibitory autapse mediates anticipated synchronization between coupled neurons}

\author{Marcel A. Pinto}
\affiliation{Instituto de F\'{\i}sica, Universidade Federal de Alagoas, Macei\'{o}, Alagoas 57072-970 Brazil}
\author{Osvaldo A. Rosso}
\affiliation{Instituto de F\'{\i}sica, Universidade Federal de Alagoas, Macei\'{o}, Alagoas 57072-970 Brazil}
\affiliation{Departamento de Inform\'atica en Salud, Hospital Italiano de Buenos Aires $\&$ CONICET,  C1199ABB, Ciudad Aut\'onoma de Buenos Aires, Argentina.}
\affiliation{Complex Systems Group, Facultad de Ingenier\'{\i}a y Ciencias Aplicadas, Universidad de los Andes, Avenida Monseñor \'Alvaro del Portillo 12.455, Las Condes, Santiago, Chile.}

\author{Fernanda S. Matias}
\thanks{fernanda@fis.ufal.br}
\affiliation{Instituto de F\'{\i}sica, Universidade Federal de Alagoas, Macei\'{o}, Alagoas 57072-970 Brazil}

\begin{abstract}
Two identical autonomous dynamical systems unidirectionally coupled in a sender-receiver configuration can exhibit anticipated
synchronization (AS) if the Receiver neuron (R) also receives a delayed negative self-feedback. Recently, AS was shown to
occur in a three-neuron motif with standard chemical synapses where the delayed inhibition was provided by an interneuron.
Here we show that a two-neuron model in the presence of an inhibitory autapse, which is a massive self-innervation present in the cortical architecture, may present AS.
The GABAergic autapse regulates the internal dynamics 
of the Receiver neuron and acts as the negative delayed self-feedback 
required by dynamical systems in order to exhibit AS.
In this biologically plausible scenario, a smooth transition from the usual delayed synchronization (DS) to AS typically occurs
when the inhibitory conductance is increased. The phenomenon is shown to be robust when model
parameters are varied within a physiological range. For extremely large values of the inhibitory autapse the system undergoes 
to a phase-drift regime in which the Receiver is faster than the Sender. 
Furthermore, we show that the inhibitory autapse promotes a faster internal dynamics of the free-running Receiver when the two neurons are uncoupled, 
which could be the mechanism underlying anticipated synchronization and the DS-AS transition.

\end{abstract}
\pacs{87.18.Sn, 87.19.ll, 87.19.lm}
\maketitle

%
%

\section{Introduction}
Flexible communication among neuronal networks requires that the same anatomical connectivity can present   
different functional connectivity patterns~\cite{Battaglia12}.
At the neuronal level, microcircuits producing coherent oscillations can be viewed as building blocks of 
the effective network dynamics in the brain.
This means that 
the structural connectivity of neuronal motifs and the intrinsic excitability of each neuron modulate the information flow in the network~\cite{Pariz18}. 
In particular, the presence of autaptic connections (or autapses), which could be synapses from the axon of a neuron to its own somato-dendritic domain,
have been shown to influence both the firing rhythm of the neuron and the synchronization properties of the network~\cite{Pawelzik03,Tamas97,Wang14,Herrmann04}.

The first observation of autapses was reported more than forty five years ago in pyramidal neurons from the neocortex~\cite{Van72}.
Since then, neurobiologists have changed the status of these connections from
not relevant structures that only appears in experimental cultures 
to a massive self-innervation present in the cortical architecture \cite{Herrmann04}. 
Inhibitory autapses can be axo-axonic, axo-dendritic or dendo-dendritic and have been found in the cerebellum~ \cite{Pouzat98,Pouzat99}, 
neocortex~\cite{Tamas97} and hippocampal cultures~\cite{Bekkers91,Bekkers02}.
However, autaptic connections are rarely included in cortical circuit diagrams 
and it is not known whether they present any kind of plasticity~\cite{Deleuze2014}.
Even though the functional role of the autapses is still lacking,
it has been speculated for decades that they could act as 
self-inhibitory function~\cite{Bekkers91}.

Here we propose that autapses could act as the negative delayed self-feedback 
required by unidirectionally coupled dynamical systems in order to present anticipated synchronization (AS)~\cite{Voss00}.
AS has been shown to be a stable but counter-intuitive solution of a system described by the following set of equations:
\begin{eqnarray}
\label{eq:voss}
\dot{\bf {S}} & = & {\bf f}({\bf S}(t)), \\
\dot{\bf {R}} & = & {\bf f}({\bf R}(t)) + K[{\bf S}(t)-{\bf R}(t-t_d)]. \nonumber 
\end{eqnarray}
${\bf f}(S)$ is a vector function that describes the autonomous dynamical
system, $K$ is the coupling matrix and the delayed term ${\bf R}(t-t_d)$ is the self-feedback~\cite{Voss00}. 
The solution $R(t) = S(t + t_d )$, characterizes
the anticipated synchronization and has been verified in a
variety of theoretical~\cite{Voss00,Voss01b,Voss01a,Kostur05,Masoller01,HernandezGarcia02,Hayashi16} and
experimental~\cite{Sivaprakasam01,Ciszak09,Tang03} studies.
The striking aspect of this solution is that in a time $t$ the Receiver $R$ predicts the state of the Sender $S$ in a future time $t+t_d$.

The first verification of anticipation in a neuronal model was done by Ciszak et al.~\cite{Ciszak03} with two FitzHugh-Nagumo neurons diffusively coupled 
in such a way that, apart from an external stimulus, they could be modeled by Eq.~\ref{eq:voss}.
AS has also been verified between two Rulkov map-based neurons in the presence of synaptic delay and a memory term as the self-feedback ~\cite{Sausedo14}, as well as 
in a three-neuron motif coupled by chemical synapses in which the self-feedback was mediated by an interneuron~\cite{Matias11}. The later result has been extended to show that AS is robust against 
noise~\cite{Matias16} and spike-timing dependent plasticity~\cite{Matias15}.
AS has also been verified in cortical-like populations in the presence of excitatory and inhibitory neurons~\cite{Matias14}, 
which could explain a positive and unidirectional Granger causality and a negative phase difference between cortical areas of a non-human primate~\cite{Matias14,Montani15,Brovelli04,Salazar12}.

Furthermore, it has been shown that the internal dynamics of the Receiver neuron may control the relative phase between the Sender and the Receiver neurons if they are synchronized~\cite{Hayashi16}.
In fact, anticipation in spike synchronization has been shown between two unidirectionally coupled nonidentical chaotic neurons when the mean frequency of the free post-synaptic neuron is greater than
the pre-synaptic one~\cite{Pyragiene13}. AS has also been verified between two Hodgkin-Huxley neurons coupled by a nonlinear excitatory synapse in which the depolarization levels 
(but not the synaptic conductance) determine the phase differences~\cite{Simonov14}. 
More recently, AS has also been verified in population neural models representing a microcircuit of sthe ongbird brain involved in song production in which the Receiver dynamics is faster than the Senders~\cite{Dima18}.

Here we show that a two-neuron motif with an inhibitory autapse in the post-synaptic neuron can present an anticipatory regime.
The autapse can act as a dynamical self-feedback and it can regulate the internal dynamics of the Receiver neuron.
We employ numerical simulations to show that GABAergic autapses can be
a biologically plausible mechanism for AS in neuronal motifs.
For very small values of the inhibitory conductance the neurons synchronize with the usual pre-post order which is called the delayed synchronization regime (DS). 
As we increase the inhibition the system presents a conductance-induced DS-AS transition.
In Sec.~\ref{model} we describe our simple motif as well as the neuronal and synaptic models.
In Sec.~\ref{results}, we report our results, showing that AS can be mediated by an inhibitory autapse in physiological regions of parameter space.
Concluding remarks and briefly discussion of the significance of our findings for neuroscience are presented in Sec.~\ref{conclusions}.

\section{\label{model}Model}

\begin{figure}
\centering
\includegraphics[width=0.8\columnwidth,clip]{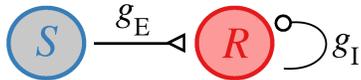}
\caption{
{\bf (Color online) Neuronal motif:} Sender (S) and Receiver (R) neurons unidirectionally connected by an excitatory chemical synapse. 
The Receiver neuron presents an inhibitory chemical autapse which is a dynamical version of the negative delayed self-feedback in Eq.~\ref{eq:voss}  .
}
\label{fig:MS}
\end{figure}

Our motif is composed of two Izhikevich neuron models~\cite{Izhikevich03} and chemical synapses as shown in Fig.\ref{fig:MS}. 
The spiking activity of the Sender (S) neuron is described by the equations bellow:
\begin{eqnarray}
\dot{v_S} &=& 0.04v_{S}^{2} + 5v_S + 140 - u_S + I \\
\dot{u_S} &=& a(bv_S - u_S).
\end{eqnarray}
The Receiver (R) neuron has similar equations but with two extra terms which accounts for the excitatory (E) and self-inhibitory (I) synaptic currents:
\begin{eqnarray}
\dot{v_R} &=& 0.04v_{R}^{2} + 5v_R + 140 - u_R + I + \nonumber  \\
          &+& g_{E}r_{E}(E_{E}-v_{S}) + g_{I}r_{I}(E_{I}-v_{R}) \\
\dot{u_R} &=& a(bv_R - u_R).
\end{eqnarray}
For both neurons the reset condition is given by: if $v_i \geq 30$~mV, then $v_i \longleftarrow c$ and $u_i \longleftarrow u_i + d$.
The variable $v_i$ is the equivalent of the membrane potential of the neuron
and $u_i$ represents a membrane recovery variable.
The subscript $i$ indexes each neuron $i={R,S}$.
We employ the same set of parameters for both neurons: $a = 0.02$, $b = 0.2$, $c = -65$ and $d = 8$.
$I$ is the external constant current which determines the neuronal firing rate.

The R neuron is subject to one excitatory synapses from the S neuron mediated by AMPA receptors with synaptic conductance $g_E$.
R is also subject to the inhibitory autapse mediated by GABA$_\text{A}$ receptors with autaptic conductance $g_I$.
The AMPA and GABA$_\text{A}$ reversal potentials are respectively 
$E_{E}=0$~mV and $E_{I}=-80$~mV.
The fraction of bound (i.e. open) synaptic receptors $r_j$ is modeled by a first-order
kinetic dynamics:
\begin{equation}
\label{eq:rate}
\dot{r}_{j} = \alpha_j [T](1-r_{j}) - \beta_j r_{j},
\end{equation}
where $\alpha_j$ and $\beta_j$ are rate constants. 
The index $j$ represents the excitatory synapse and the inhibitory autapse $j={E,I}$.
$[T]$ is the
neurotransmitter concentration in the synaptic cleft.
In its simplest model 
it is an instantaneous function of the pre-synaptic
potential $v_{pre}$~\cite{KochSegev}:
\begin{equation}
[T](v_{pre}) = \frac{T_{max}}{1+e^{[-(v_{pre}-V_p)/K_p]}}.
\end{equation}
In our model $T_{max}=1$~mM$^{-1}$, $K_p=5$~mV, $V_p=2$~mV.

Unless otherwise stated, the rate constants are $\alpha_{E}=1.1$~mM$^{-1}$ms$^{-1}$, $\beta_{E}=0.19$~ms$^{-1}$ ,
$\alpha_{I}=5.0$~mM$^{-1}$ms$^{-1}$, and $\beta_{I}=0.30$~ms$^{-1}$ similarly to the ones in Refs.~\cite{KochSegev,Matias11}.
However, these values depend on a number of different factors and can vary
significantly~\cite{Geiger97,Kraushaar00}.

%

\begin{figure}[!h]
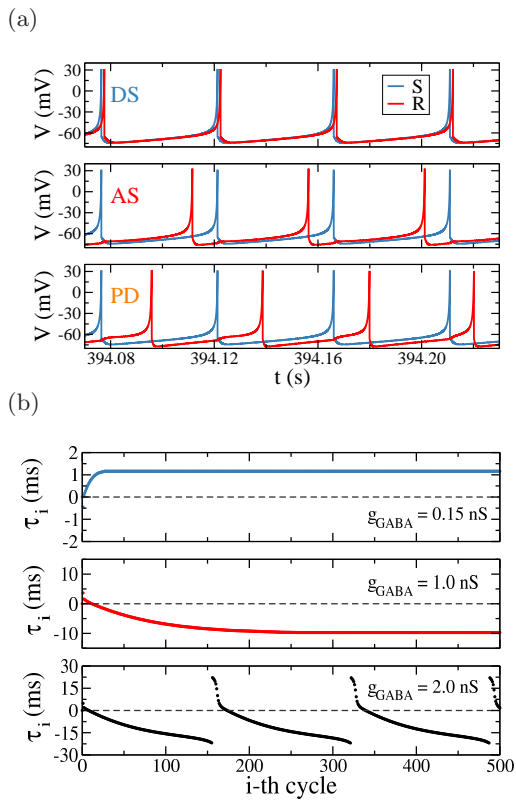
%
\begin{minipage}{0.8\linewidth}
\begin{flushleft}(a)%
\end{flushleft}%
\centerline{\includegraphics[width=0.9\columnwidth,clip]{Matias02a}}
\end{minipage}
\begin{minipage}{0.8\linewidth}
\begin{flushleft}(b)%
\end{flushleft}%
\centerline{\includegraphics[width=0.95\columnwidth,clip]{Matias02b}}
\end{minipage}
\caption{
{\bf (Color online) Characterizing the phase-locking and phase-drift regimes. } 
(a) Time series of the Sender and the Receiver neuron after a transient time for different values of the autaptic inhibition $g_I$ and fixed external current $I=10$ pA. 
(b) The spike-timing difference $\tau_i$ between S and R as a function of the i-th cycle. 
A pre-post firing order characterizes the usual delayed (DS) synchronization regime (upper panels,  $g_I=0.15$~nS). 
The anticipated synchronization regime (AS) occurs when the Receiver neuron fires before the Sender (middle panels,  $g_I=1.0$~nS).
When neurons fire with different frequencies, the spike timing difference changes every cycle and the neurons are in a phase-drift (PD) regime (bottom panels,  $g_I=2.0$~nS).
}
\label{fig:Vmemb}
\end{figure}

\section{\label{results}Results}

Initially, we describe our results for the scenario where both neurons
are subjected to a constant current $I\geqslant 5$~pA. 
Unless otherwise stated we keep the excitatory conductance fixed as $g_E=0.3$~nS 
which is larger enough to promote a phase-locking regime between the two neurons when there is no autapse. 
For different sets of inhibitory
conductance values $g_I$ our motif can exhibit different behaviors as shown in Fig.~\ref{fig:Vmemb}. In order to
characterize each regime, we define $t^{S}_{i}$ as the time in which the
membrane potential of the Sender neuron is larger than the threshold ($v_i \geq 30$~mV) in the $i$-th
cycle (i.e. its $i$-th spike time), and $t_{i}^{R}$ as the spike time
of the Receiver neuron which is nearest to $t_i^{M}$.
The spike-timing difference $\tau$ between S and R  is defined as:
\begin{equation}
\tau_i \equiv t^{R}_{i}-t^{S}_{i}.
\end{equation}

When $\tau_i$ converges to a constant value $\tau$, a
phase-locked regime is reached~\cite{Strogatz}. By definition, if $\tau > 0$
we say that the system exhibits delayed
synchronization (DS) and the neurons fire in the usual pre-post order (see upper panels in Fig.~\ref{fig:Vmemb}a and b, for $g_I=0.15$~nS) . 
If $\tau<0$  we say that anticipated synchronization (AS)
occurs and the neurons fire in a non-intuitive post-pre order (see middle panels in Fig.~\ref{fig:Vmemb}a and b, for $g_I=1.0$~nS). If $\tau_i$ does not converge to a fixed
value, the system is in a phase-drift (PD) regime~\cite{Strogatz}~(see bottom panels Fig.~\ref{fig:Vmemb}a and b, for $g_I=2.0$~nS).

\begin{figure}[!h]
\centering
\includegraphics[width=0.8\columnwidth,clip]{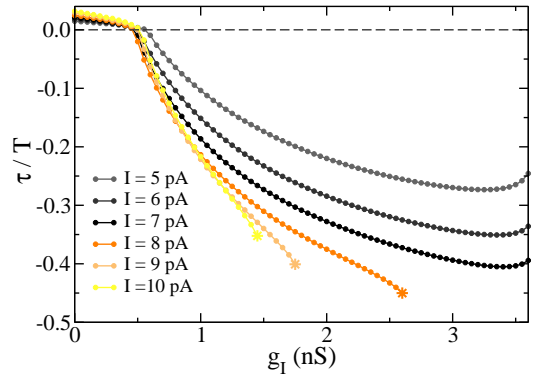}
\caption{
{\bf (Color online) Transition from delayed to anticipated synchronization regime. } 
Spiking time difference normalized by the period of the phase-locking $\tau/T$ as a function of inhibitory conductance $g_I$ for different values of external current $I$. 
Positive values of $\tau$ characterizes DS, whereas $\tau<0$ represents the AS regime. 
The stars mark the end of the phase-locking regime and the beginning of the phase-drift (PD). 
In PD the value of $\tau_{i}$ in each cycle does not converge to a fixed value $\tau$ as shown in Fig.~\ref{fig:Vmemb}(b).
}
\label{fig:tau_gis}
\end{figure}

\subsection{\label{sec:MS} The conductance induced transition from DS to AS}

\begin{figure}[!ht]
\centering
\includegraphics[width=0.99\columnwidth,clip]{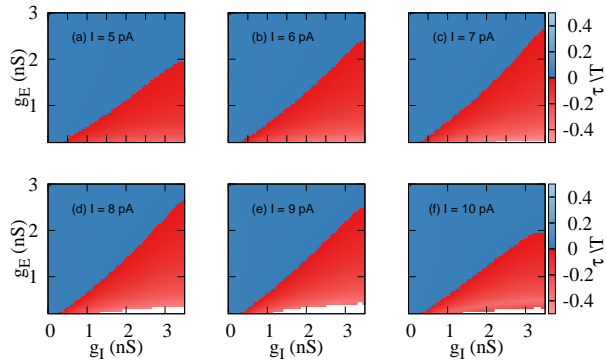}
\caption{
{\bf (Color online) Normalized spike-timing difference $\tau/T$ (right bar) in the ($g_I$,$g_E$) projection of parameter space} 
for different values of external current $I$: DS (blue, mostly at the upper part in which $g_I>g_E$ and $\tau/T<0$), AS (red, middle), and PD
(white, meaning that no stationary value of $\tau$ was found).}
\label{fig:ge_gi_tau}
\end{figure}

The spike-timing difference $\tau$ is a continuous and smooth function of $g_{I}$.
The transition from  delayed to anticipated regime through a zero-lag regime 
may be induced by the inhibitory autaptic conductance as shown in
Fig.~\ref{fig:tau_gis}. To compare the transition for different values of external current we normalized $\tau$ by the firing period $T$ which depends on $I$. 
Both AS and the DS-AS transition can be found in a large region of parameter space. Typically, AS occurs for $g_I>g_E$.
The relation between $\tau/T$, excitatory and inhibitory conductances for different values of $I$ is shown in Fig.~\ref{fig:ge_gi_tau}.
Results are independent of initial conditions and perturbations. The phase-locking regimes are always reached after a transient time.

It is worth to mention that the presence of the autapse in the post-synaptic neuron, even for very small values of inhibitory conductance, 
decreases the spike-timing difference between the neurons. This means that autapses could be biologically important to overcome synaptic delays between distant areas. 
Moreover, for enough inhibition, the neurons can fire exactly at the same time. 
Therefore, autapse can be also considered as a mechanism to promote zero-lag synchronization\cite{Gray89,Gollo14} which does not require bidirectional connections.

We show here that autaptic weights may control the spike-timing difference between pre and post-synaptic neurons through a conductance induced DS-AS transition. 
This result can be computationally considered as the other way around of the spike-timing dependent plasticity (STDP)~\cite{Abbott00}, which is 
a biological mechanism that allows the spike-timing differences between pre and post-synaptic neurons to control changes in the synaptic weights.

\subsection{\label{sec:faster} GABAergic autapse promotes a faster internal dynamics}

\begin{figure}[!h]%
\begin{minipage}{0.8\linewidth}
\begin{flushleft}(a)%
\end{flushleft}%
\centerline{\includegraphics[width=0.9\columnwidth,clip]{Matias05a}}
\end{minipage}
\begin{minipage}{0.8\linewidth}
\begin{flushleft}(b)%
\end{flushleft}%
\centerline{\includegraphics[width=0.9\columnwidth,clip]{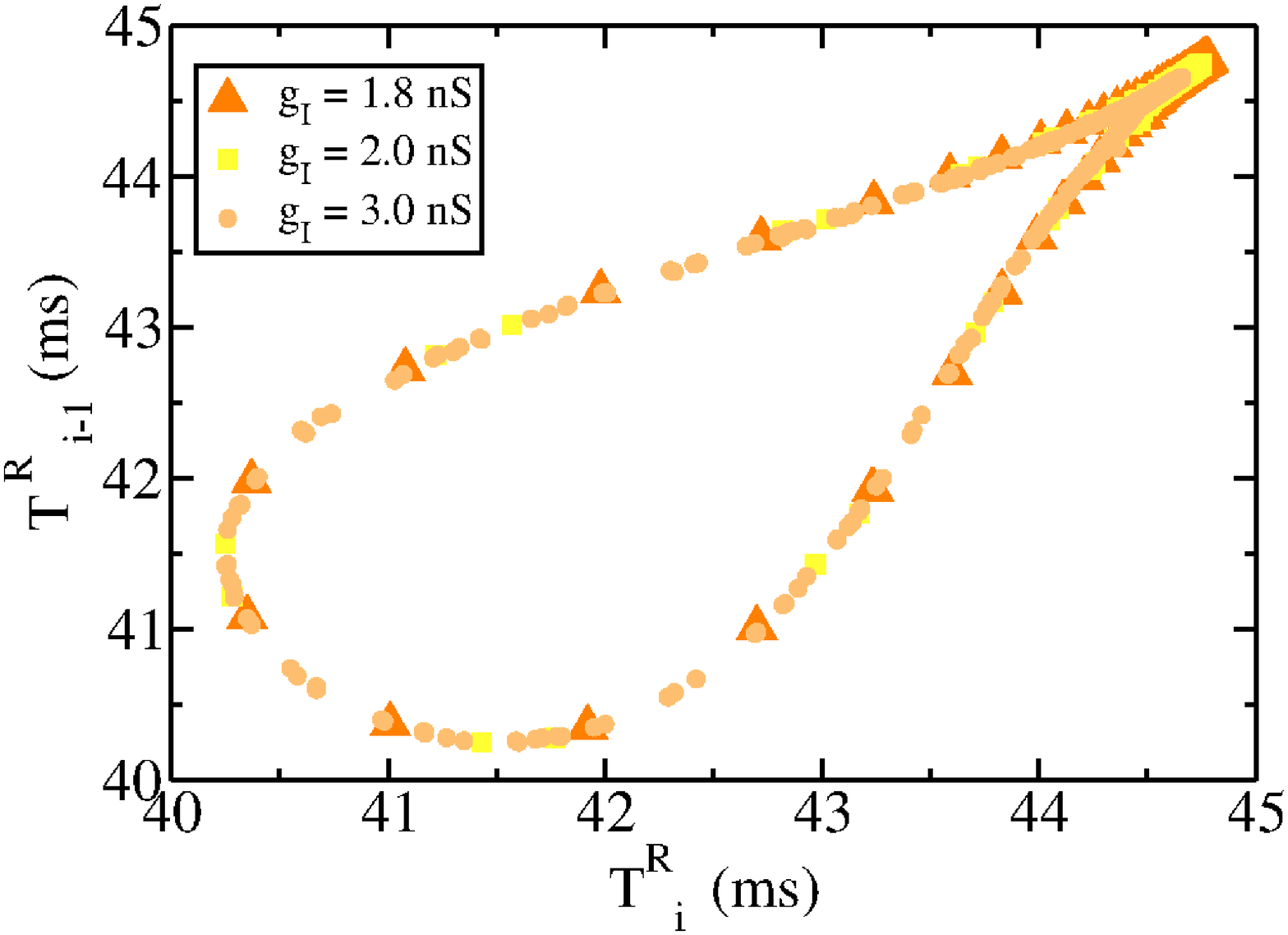}}
\end{minipage}
\caption{
{\bf (Color online) Characterizing the phase-drift regime induced by the inhibitory autapse: the Receiver is faster than the Sender.} 
(a) The mean period of the Receiver normalized by its own period when $g_{I}=0$, $\overline{T_{R}}/T_{0}$
coincides with the mean period of the Sender for DS and
AS regimes, but it is smaller for PD. 
(b) In PD, the return map of
the period in each cycle of the Receiver is consistent with a quasi-periodic
system.}
\label{fig:returnmap}
\end{figure}

Occasionally inhibitory synapses are viewed as a biological mechanism to avoid neurons to fire. 
However, we show here that the GABAergic autapse can allow the neuron to fire with higher frequencies.
In the phase-locking regimes (both DS and AS) the period of the Receiver is the same of the Sender ($T_{R}=\overline{T_{R}}=T_0$ which is the period of a neuron without an autapse $g_I=0$).
As we increase the inhibition, the spike-timing difference decreases and eventually, the system loses the phase-locking regime. 
For $g_E=0.3$ nS and $I\geq8$~pA, as we increase $g_I$ the system exhibits a second transition from AS to a phase-drift regime in which $\overline{T_{R}}<T_0$ (see Fig.~\ref{fig:returnmap}(a)). 
This means that in the PD regime the Receiver fires faster than the Sender. 
The return map of the Receiver period in each cycle $T^{R}_{i-1}$ versus $T^{R}_{i}$ characterizes the phase-drift (see Fig.~\ref{fig:returnmap}(b)) which is a quasi-periodic regime.
This counter-intuitive effect of inhibition has also been reported for a neuron participating in a inhibitory loop mediated by an interneuron~\cite{Matias11}.

For small values of external current $I<8$~pA and $g_I>3.6$~nS
the membrane potential of the Receiver neuron does not reach the condition for reset, which means that it does not fire a spike. Thus, it is not possible to define $\tau_i$. 
In fact, for this region of the parameter space the inhibition can silence the neuron.

\begin{figure}[!ht]
\centering
\includegraphics[width=0.8\columnwidth,clip]{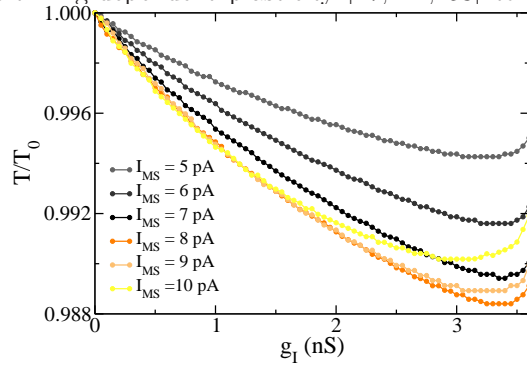}
\caption{
{\bf (Color online) Inhibitory autapse promotes a faster internal dynamical of the free-running Receiver,} 
 which could be the mechanism promoting AS and the transition from DS to AS. 
 For the uncoupled situation ($g_{E}=0$) the period of the free-running Receiver (normalized by its own period for $g_{I}0$, $T_{R}/T_{0}$) decreases as we increase the inhibitory conductance $g_{I}$.
 For external currents $I=5,6,7$~pA the minimum value of $T/T_o$ can also be related to the minimum value of $\tau$ in Fig.~\ref{fig:tau_gis}. 
 For $I\geq 8$~pA the system undergoes a transition from AS to PD regime before $T_{R}/T_{0}$ reaches the minima. 
  }
\label{fig:uncoupled}
\end{figure}

The effect of the autapse in the internal dynamics of the Receiver can also be studied for the uncoupled case $g_E=0$. 
For this situation, as we increase $g_I$, the firing period of the neuron subjected to the autapse initially decreases and $T/T_0<1$ for any value of external current $I$ (see Fig.~\ref{fig:uncoupled}). 
This result is in agreement with previous studies showing 
that the mechanism underlying anticipated synchronization is a faster internal dynamics of the Receiver~\cite{Hayashi16,Pyragiene13,DallaPorta19}.

The inhibitory autapse can also promote the transition from DS to AS when the sender and the receiver neurons are not described by the same equations or parameters.
If the receiver neuron is a fast spiking neuron~\cite{Izhikevich03} 
(with model parameters $a = 0.01$, $b = 0.2$, $c = -65$ and $d = 2$) the transition from positive to negative $\tau$ is still continuous and smooth.
In such case the external current in each neuron should be different in order to ensure that their natural frequencies are similar. 
For example, if $I_S = 16$~pA, $I_R = 4.5$~pA, $g_E = 0.3$~nS, $\beta_I = 0.188$~ms$^{-1}$, $\beta_E=0.3$~ms$^{-1}$ 
the system exhibits DS if $g_I\lesssim 0.1$ and AS if $0.1 \lesssim g_I \lesssim 0.5$~nS (data not shown).

Moreover, if the receiver neuron is modeled as a two-compartment neuron, the system can also exhibit AS. 
We have considered that the compartments are connected by an electrical synpase
with synaptic current given by $I = g_{electrical}*(V_{pre} - V_{post})$ and the second compartment sends an inhibitory chemical synapse to the first. 
For $g_E=1$~nS, $g_{electrical}=0.2$~nS, $I=10$~pA the system undergoes a transition from DS to AS when we increase  $g_I$ from zero to $2.5$~nS (data not shown).
Further investigation on how anticipated synchronization depends on more complex dendritic trees, 
other types of neurons and slower synaptic currents mediated by NMDA and GABA$_\text{B}$
are beyond the scope of this paper and should be reported elsewhere in the future.

%
%
%

\section{\label{conclusions}Concluding remarks}

To summarize, we have shown that the presence of a GABAergic autapse can promote anticipated synchronization between two coupled neurons.
Here, the inhibitory self-innervation acts as a dynamical delayed negative self-feedback~\cite{Voss00,Ciszak03} 
or the inhibitory loop mediated by a third neuron~\cite{Matias11,Matias14} required by previously studied systems in order to exhibit AS. 
This also means that even in a very simple microcircuit of two neurons, the characteristic time scale and conductance of the excitatory connection are not enough to
determine the phase difference between two synchronized neurons. 
Moreover, autapses could be a biologically plausible mechanism to overcome synaptic delays. 

We have also shown that the neuron subjected to the inhibitory autapse can fire faster than without it.
Such a result is in consonance with previous studies~\cite{Hayashi16,Pyragiene13,DallaPorta19} showing that the mechanism underlying 
AS and the DS-AS transition can be a faster internal dynamics of the Receiver system.
Moreover, previous work on AS have investigate the phenomenon for neurons with firing rates larger than $60$~Hz~\cite{Matias11,Matias15}.
Typically cortical neurons in vivo do not fire at rates greater than $\sim 30$~Hz. 
Here we show that AS can occur for firing rates around $20$~Hz which means that AS is a robust phenomenon at different time scales.

The DS-AS transition could possibly explain commonly reported short latency in visual systems~\cite{Orban85,Nowak95,Kerzel03,Jancke04,Puccini07,Martinez14}, olfactory circuits~\cite{Rospars14}, 
songbirds brain~\cite{Dima18} and human perception~\cite{Stepp10,Stepp17}.
Differently from the first papers about AS~\cite{Voss00,Ciszak03}, here the anticipation time is not hard-wired in the dynamical equations, but rather emerges from the autapse dynamics.
The spike-timing difference between post- and pre-synaptic neurons decrease as we increase the inhibitory autaptic conductance. 
Eventually the system presents a conductance induced transition from delayed to anticipated regimes similarly to what has been reported in a three neuron motif~\cite{Matias11,Matias14}.
This AS-DS transition could synergistically work together with spike-timing-dependent
plasticity~\cite{Abbott00,Clopath10,Matias15} to determine the circuit dynamics and synaptic conductances. Including effects from synaptic plasticity 
in the model presented here is a natural next step which we are
currently pursuing.


\begin{acknowledgments}
The authors thank FAPEAL and UFAL (for PIBIC grant), CNPq (grant 432429/2016-6) and CAPES (grant 88881.120309/2016-01) for finnancial support.

\end{acknowledgments}
%

\bibliography{matias}

\end{document}